\documentclass[sigconf]{acmart}

\AtBeginDocument{%
  \providecommand\BibTeX{{%
    \normalfont B\kern-0.5em{\scshape i\kern-0.25em b}\kern-0.8em\TeX}}}
\usepackage{multirow}
\usepackage{enumitem}
\usepackage{color,soul}

\copyrightyear{2025}
\acmYear{2025}
\setcopyright{cc}
\setcctype{by-sa}
\acmConference[CHI '25]{CHI Conference on Human Factors in Computing Systems}{April 26-May 1, 2025}{Yokohama, Japan}
\acmBooktitle{CHI Conference on Human Factors in Computing Systems (CHI '25), April 26-May 1, 2025, Yokohama, Japan}\acmDOI{10.1145/3706598.3713088}
\acmISBN{979-8-4007-1394-1/25/04}

\begin{document}

\title[What Did My Car Say? Impact of Autonomous Vehicle Explanation Errors and Driving Context]{What Did My Car Say? Impact of Autonomous Vehicle Explanation Errors and Driving Context On Comfort, Reliance, Satisfaction, and Driving Confidence}

\author{Robert Kaufman}
\email{rokaufma@ucsd.edu}
\affiliation{%
  \institution{University of California, San Diego}
  \streetaddress{9500 Gilman Dr}
  \city{La Jolla}
  \state{California}
  \country{USA}
  \postcode{92093}
}

\author{Aaron Broukhim}
\email{aabroukh@ucsd.edu}
\affiliation{%
  \institution{University of California, San Diego}
  \city{La Jolla}
  \state{California}
  \country{USA}
}

\author{David Kirsh}
\email{kirsh@ucsd.edu}
\affiliation{%
  \institution{University of California, San Diego}
  \city{La Jolla}
  \state{California}
  \country{USA}
}

\author{Nadir Weibel}
\email{weibel@ucsd.edu}
\affiliation{%
  \institution{University of California, San Diego}
  \city{La Jolla}
  \state{California}
  \country{USA}
}

\renewcommand{\shortauthors}{Kaufman et al.}

\begin{abstract} 
Explanations for autonomous vehicle (AV) decisions may build trust, however, explanations can contain errors. In a simulated driving study (n = 232), we tested how AV explanation errors, driving context characteristics (perceived harm and driving difficulty), and personal traits (prior trust and expertise) affected a passenger's comfort in relying on an AV, preference for control, confidence in the AV's ability, and explanation satisfaction. Errors negatively affected all outcomes. Surprisingly, despite identical driving, explanation errors reduced ratings of the AV's driving ability. Severity and potential harm amplified the negative impact of errors. Contextual harm and driving difficulty directly impacted outcome ratings and influenced the relationship between errors and outcomes. Prior trust and expertise were positively associated with outcome ratings. Results emphasize the need for accurate, contextually adaptive, and personalized AV explanations to foster trust, reliance, satisfaction, and confidence. We conclude with design, research, and deployment recommendations for trustworthy AV explanation systems.
\end{abstract}

\ccsdesc[500]{Human-centered computing~Empirical studies in HCI}
\ccsdesc[500]{Human-centered computing~User studies}
\keywords{Autonomous Vehicles, Explainable AI, AI Errors}


\maketitle

\section{Introduction}
Autonomous vehicles offer a wide range of potential societal benefits, including reduced driving infractions, traffic volume, environmental impact, and passenger stress~\cite{fagnant2015preparing}. Despite these benefits, it is a well-documented problem that adoption is limited by a lack of trust in how vehicles make decisions~\cite{kenesei2022trust}. This is not a problem specifically with autonomous vehicles, but a widespread issue across many types of AI-based systems~\cite{bedue2022can}. System transparency via explainable AI (XAI) has been proposed as a means to mitigate concerns with AI-based systems like AVs, offering users a look “under the hood” of black-box AI models so they understand what the system is doing and why~\cite{gunning2019darpa, miller2019explanation}. 

However, although potentially increasing trust, explanation of AI \color{black} behavior and decision-making \color{black} in the real world can contain errors.  Prior work has shown that when autonomous vehicles exhibit \emph{driving} errors, passenger trust and willingness to rely on the vehicle can deteriorate quickly~\cite{seet2020differential, kaplan2023trust}. Far less is known about the impact of \emph{explanation} errors\color{black}, such as when an AV miscommunicates what it is doing or why\color{black}~\cite{cabitza2024explanations, kenny2021explaining}. Particularly for safety-critical systems like autonomous vehicles -- that may rely on explanations and other in-vehicle communications to elicit trust comfort, and safe reliance with users -- knowing the consequences of errors is pivotal to safe deployment. \color{black} We hypothesize that explanation errors, like driving errors, will negatively impact perceptions of an AV. \color{black} Understanding these consequences is essential for user reliance because, without this knowledge, AV systems cannot be deployed safely or ethically~\cite{martinho2021ethical}.

\color{black} Autonomous vehicles offer a unique context to study explainable AI errors, as vehicle errors may be easier to detect for lay users with basic driving knowledge compared to errors in more specialized domains, such as medical diagnosis \cite{rajpurkar2017chexnet, kaufman2023explainable} or bird identification \cite{pazzani2022expert, soltani2022user}, where domain expertise is required for error detection. This distinct characteristic allows non-experts to evaluate AV explanations with relatively high confidence, providing a unique perspective on how explanation errors might influence user perceptions, judgments, and trust in AI. Prior research in XAI suggests that task complexity \cite{salimzadeh2023missing} and domain expertise \cite{nourani2020role} are crucial factors in how users interpret and respond to AI explanations -- we extend this understanding by examining explanation errors in a high-risk, non-expert domain, contributing new insights into how explanation accuracy affects user trust and reliance. \color{black}

Human-AV interactions do not exist in a vacuum: they are sensitive to the contextual demands of the external driving environment~\cite{hoff2015trust}. It is important to consider the driving context in which explanation errors occur, as this may significantly influence how people interact with AI-based systems~\cite{lim2009and, schilit1994context}, including AVs~\cite{capallera2022human, de2020designing}. The complexity of a driving situation and perceived risk of harm have been singled out as particular factors of interest~\cite{ha2020effects, kaufman2024developing}, as these might drive a person's explanatory needs and reliance behaviors. \color{black} Though presently underexplored, we hypothesize that contextual factors may impact rider perceptions of an AV, particularly when errors are introduced. \color{black}

Finally, recent work has posited that AV explanations should be tailored to meet the specific needs of the people interacting with the system~\cite{ma2023analysing, kaufman2024developing}. It is well known that people may interact differently with AI-based systems~\cite{schneider2019personalized} based on the individual characteristics or prior experiences they may have~\cite{ayoub2021modeling, kaufman2024predicting}. Particularly well studied are domain expertise~\cite{araujo2020ai, pazzani2022expert, kaufman2022cognitive} and initial (dispositional) trust~\cite{hoff2015trust}; \color{black} we hypothesize that these will also impact AV perceptions.\color{black}

In the present study, we examine the impact of explanation errors across a variety of realistic, simulated driving scenarios. To deepen our investigation and understand the significance of the \textit{type} of error presented, participants were shown AV explanations at three distinct accuracy levels: \color{black}(1) accurate explanation of the AV's behavior, (2) accurate explanation of \textit{what} the AV is doing but incorrect rationale for \textit{why}, and (3) incorrect explanation of what \textit{and} rationale of why. \color{black} We measured the effect of these errors on four main outcomes: comfort relying on the AV, preference for control, confidence in the AV's driving ability, and explanation satisfaction. We include measures of scenario context -- perceived harm and driving difficulty -- to assess their impact on our driving outcomes, including how they may moderate the relationship between explanation errors and user perceptions. To explore the influence of individual differences, we also included measures of trust and \color{black}AV domain \color{black}expertise to determine if they predict study outcomes.

Our findings reveal that explanation errors, contextual characteristics, and personal traits significantly impact how a person may think, feel, and behave towards AVs. Explanation errors negatively affected all outcomes, with impacts proportional to the magnitude and negative implications of the error. Harm and driving difficulty directly impacted outcomes as well as moderated the relationship between errors and outcomes, though in opposing ways. Overall, harm was generally seen as more important and more negative than difficulty. Participants with higher \color{black}AV domain \color{black}expertise tended to trust AVs more, and these each correlated with more positive outcome ratings in turn.

\vspace{1em}
\noindent
In sum, \color{black} our research questions are as follows:
\begin{itemize}[topsep=0pt, nolistsep]
    \item \textbf{RQ1:} What is the impact of AV explanation errors on participant ratings of an AV, and does the impact differ for errors of different information types (`why' errors vs. `what and why' errors)?
    \item \textbf{RQ2:} What is the impact of context (perceived harm and driving difficulty) on participant ratings of an AV, and do these contextual factors modify the effect of an error?
    \item \textbf{RQ3:} How does participant initial trust and domain expertise impact their ratings of an AV?
    \item \textbf{RQ4:} How can findings inform future design and research of explainable AI and autonomous vehicles?
\end{itemize}
\color{black}
\vspace{1em}
\noindent
Results highlight the critical need for accurate and contextually adaptive explanations for autonomous vehicles to enhance user trust, reliance, satisfaction, and confidence. Recognizing the implications of explanation errors is vital for advancing AV research and guiding design teams to make informed decisions. This understanding lays the groundwork for developing context-aware designs, personalized explanation interfaces, and establishing ethical or regulatory guidelines to ensure the deployment of safe and trustworthy explainable AI (XAI) systems for autonomous vehicles.

\section{Related Work}
\subsection{AV Trust and Explainability}
\textbf{Human-Centered Explainable AI} -- Continuous development of explainable AI (XAI) has brought major progress in the quest for trust through AI transparency~\cite{miller2019explanation}. Approaches to XAI vary, often limited by the availability of the model~\cite{simonyan2013deep, ribeiro2016should}. Recent work found that -- even when an explanation is given -- trust and engagement with the system may not improve unless the \emph{right information} is given in the \emph{right way}, at the \emph{right time}~\cite{wang2019designing, liao2021human}. Several studies and theory pieces have demonstrated the value of human-interpretable explanations on user understanding and trust ~\cite{holzinger2019causability, soltani2022user}. In particular, explanations that are modeled off those given by human experts have suggested as a means to increase understanding for experts and novices alike~\cite{pazzani2022expert, kaufman2022cognitive}. There have likewise been calls for the need for explanations to be sensitive to particular user characteristics (such as personal traits or experiences) ~\cite{ehsan2020human, ehsan2021explainable, kaufman2023explainable} or context of use (including the specific use environment and goals of a user)~\cite{schilit1994context, kaufman2024developing}. Despite the large amount of research on explanation design, little has been done to understand what happens when these explanations fail. The study we present here is a first step towards filling this knowledge gap, using autonomous vehicles as a specific domain of interest.

\textbf{Interface Modalities} -- Research on in-vehicle interfaces for explanation, such as visual heads-up-displays (HUDs)~\cite{currano2021little, chang2016don, schartmuller2019text}, audio interaction~\cite{mok2015understanding, jeon2009enhanced, locken2017towards}, and even haptic feedback for drivers~\cite{di2020haptic} has shown that there are a variety of ways explanations can support users, each dependent on the use-case and context of interest. The benefits and drawbacks of different modalities often relate to the complex process of transferring sufficient knowledge to accomplish a task (such as to build understanding through system transparency) without creating too much cognitive load~\cite{kaufman2024effects, kim2023and, colley2021effects}. In the present study, we leverage video and audio explanations for our study on the impact of errors, as these are common and efficient methods to transfer information to a user without overwhelm. We present both modalities of information at the same time to increase the accessibility of our study and ensure there is successful transfer of information.

\textbf{Explanation Content} -- Choosing the appropriate content for an AI explanation is crucial for enhancing rider trust and reliance \cite{miller2019explanation}. Recent work as focused on providing a description of \emph{what} a vehicle is doing and \emph{why} a vehicle is doing it, as these enable a user to create a momentary evaluation of the AV's behavior in terms of reliance \cite{hoff2015trust}. For example,~\citet{koo2015did} present `how', `why', and a combination of both in various simulated autonomous driving scenarios to assess driver attitudes and safety performance. They found improved safety when both 'how' and `why' were presented to drivers, but a preference for `why' explanations alone. \citet{kaufman2024effects} leveraged auditory and visual explanations to teach humans to be better drivers via an 'AI coach', highlighting the additive value of \emph{what} and \emph{why}-type information for transferring knowledge, adding that \textit{too much} information can cause overwhelm. They emphasize the need for explanations to strike a balance between complexity and comprehensiveness. The impact of \emph{what} (similar to Koo's \emph{how}) and \emph{why}-type explanation \emph{errors} -- explored in the present study -- remains unknown.

\textbf{Factors Impacting Trust} -- Knowing what factors may impact a person's trust and reliance decisions is vitally important to designing XAI explanations to support them. Hoff and Bashir's theoretical model of trust in automation highlights the importance of three distinct yet interdependent facets of trust: dispositional (e.g. personality or cultural attitudes), situational (e.g. based on a particular context of use), and learned trust (e.g. based on a present evaluation of system performance) \cite{hoff2015trust}. Additionally,~\citet{kaufman2024developing} developed a framework to understand situational awareness in joint action between humans and AV, a key focus of explainable AI transparency in safety-critical situations like driving. They describe how communications like AV explanations can enable human-AV teamwork to achieve particular goals like safe and trustworthy driving. Factors of interest include external driving conditions, human traits and abilities, and communication preferences and goals -- all of which are crucial in managing driving difficulty and reducing the risk of harm. Using these system-based models as a backdrop, AI explanation designers can form hypotheses on how to build more trustworthy systems. In the present study, we build off these models by investigating specific driving context factors \color{black}(perceived harm and driving difficulty) \color{black} and personal traits \color{black}(domain expertise and prior trust) \color{black}which may impact a person's reliance judgments and, in the case of context, how much an error matters.

\textbf{Knowledge Gap} -- Indeed, explanation interfaces have been implemented in mainstream deployed autonomous vehicles, such as communication interfaces by Tesla~\cite{TeslaModelYManual} and Waymo~\cite{WaymoOne}. Despite these efforts, we still know very little on the impact of AV explanation errors on how a person will trust and interact with an AV. We know even less about how errors may depend on contextual factors like driving difficulty or harm~\cite{capallera2022human, de2020designing, ha2020effects}. With this work, we seek to understand how contextual harm and difficulty of driving may affect study outcomes, as well as contribute to the body of literature on how personal traits predict a person's interactions with an AV.

\subsection{AI Errors}
Widespread integration of AI systems into everyday tasks has demonstrated huge benefits to productivity and optimization in many domains~\cite{fauzi2023analysing}. However, concerns over AI errors remain a major point of contention, particularly as systems proliferate. Examples of errors with real-world consequences include language models' propensity to hallucinate~\cite{xu2024hallucination} and algorithmic bias that favors men over women used in the hiring process~\cite{dastin2018amazon}. \color{black} In contrast to fields requiring domain-specific knowledge for error detection (e.g., medical diagnosis \cite{rajpurkar2017chexnet, kaufman2023explainable}), AV explanation errors are often intuitively detectable by lay users. This accessibility may amplify the impact of errors on trust and reliance, as errors are not only observed but judged against personal driving ability. \color{black}

\textbf{General AV Errors} -- Autonomous vehicle driving errors have important real-world consequences, which may include physical harm or even fatalities~\cite{favaro2017examining}. Even in cases where AV driving outperforms humans~\cite{schwall2020waymo}, concerns over vehicle errors can be a major hindrance to adoption and use~\cite{choi2015investigating}. \citet{luo2020trust} shows that errors caused by an AV had a more significant negative impact on user trust than external errors, such as those caused by other drivers or road conditions. Declines in trust may be difficult to recover from and have long-lasting effect \cite{seet2020differential}. Explanations are no panacea: trust is difficult to achieve when the system itself performs poorly, even when explanations meant to elicit trust are presented~\cite{kaplan2023trust}. In this paper, we seek to connect prior work showing the dire impact of driving errors \cite{luo2020trust, zhang2022trust} to broader XAI literature investigating user perceptions of AI explanations \cite{lebovitz2019diagnostic, cabitza2024explanations}.

\textbf{Explanation Errors} -- Some prior work has investigated the impact of \emph{explanation} errors for autonomous systems. In a study on “white box” XAI,~\citet{cabitza2024explanations}  found that non-expert users tend to not catch explanations errors and believe the system even when explanations were wrong, attributing the phenomena to the Halo Effect found in social psychology where people assumed correctness of the system without verifying accuracy. Over- or under-reliance is a problem as people learn to calibrate their interactions with AI systems~\cite{endsley2018situation}. Conflicting results suggest that explanations may help reduce over-reliance in some cases~\cite{vasconcelos2023explanations}, but increase over-reliance in others~\cite{kenny2021explaining}. Other research has shown that the influence of explanations on reliance may be based on the systems' performance itself~\cite{papenmeier2022s}.

\textbf{Knowledge Gap} -- Though several prior studies have investigated the consequences of accurate AV explanations, the impact of explanation errors on reliance behavior and related outcomes remains unexplored. Of particular interest are cases when the autonomous vehicle's driving performs properly, as this allows us to separate the impact of driving performance from the impact of explanation performance. We address this knowledge gap.

\begin{figure*}[t] 
  \centering  \includegraphics[width=\linewidth]{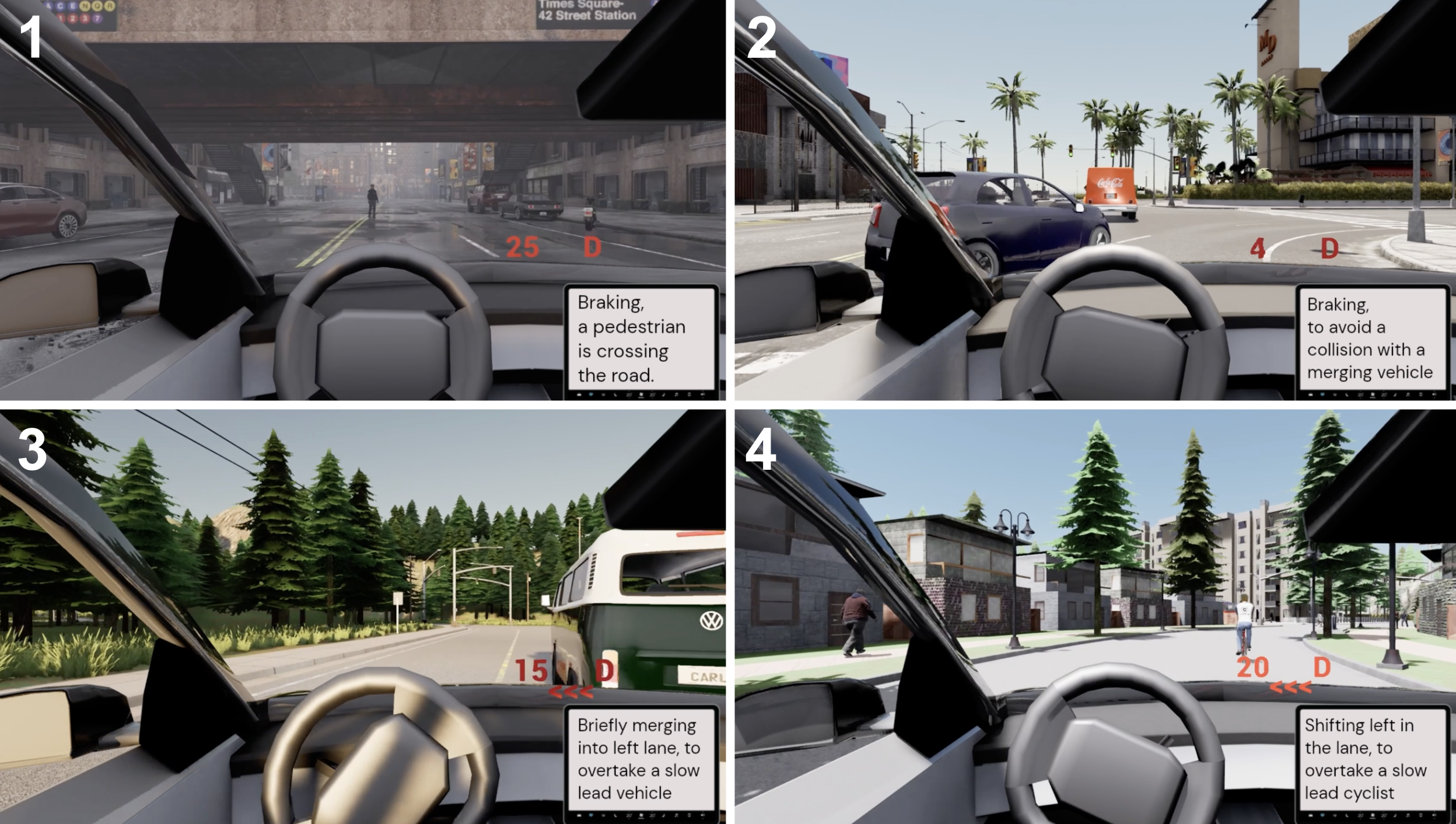}
  \vspace{-.5em}
  \caption{Example images from four driving scenarios: (1) the AV slows for a pedestrian crossing the road on a foggy day in the city, (2) the AV is cut off by a vehicle turning right from the left lane, (3) the AV merges around a slow lead vehicle in a forested area, (4) the AV moves past a cyclist on a suburban street. Written AV explanations appear on the vehicle's dashboard display.}
  \Description{Example images from four driving scenarios: (1) the AV slows for a pedestrian crossing the road on a foggy day in the city, (2) the AV is cut off by a vehicle turning right from the left lane, (3) the AV merges around a slow lead vehicle in a forested area, (4) the AV moves past a cyclist on a suburban street. Written AV explanations appear on the vehicle's dashboard display.}
  \label{scenarioexamples}
\end{figure*}

\section{Method}
We conducted an online experiment using realistic, simulated driving scenarios of an AV driving through various environments. These ranged from rural to urban driving and from routine navigational challenges like driving around construction cones to challenging situations like avoiding collisions with erratic drivers. Participants viewed videos of the scenarios and, after each, provided ratings related to trust, reliance, satisfaction, and evaluation of the AV.

To test the impact of explanation errors, participants were shown three versions of each scenario, \color{black}each differing by the accuracy of the information. This within-subjects design helps us control for individual differences and attribute findings to the error conditions and scenarios themselves. \color{black} In all three versions of each scenario, the AV drove identically; only the explanations differed. The AV always drove accurately and lawfully. Participants were randomly presented 9 of 27 possible scenarios, \color{black} resulting in a unique set for each. This approach helped evenly distribute any order effects across participants, reducing systematic bias that could result from a block design, increasing generalizability while preserving a naturalistic flow. \color{black} Each participant provided 27 total video ratings (9 scenarios X 3 accuracy levels). Scenarios were presented in random order; \color{black} all had approximately the same number of total ratings. \color{black} Scenarios are described in Table \ref{scenario_list}.

\begin{table*}[h] 
  \caption{Scenario \color{black}Reference Codes and \color{black}Descriptions}
  \Description{A list of all of the Scenarios used in the study, including brief descriptions of each.}
  \label{scenario_list}
  \vspace{-1em}
  \renewcommand{\arraystretch}{1}
  \begin{tabular}{p{0.05\linewidth}p{0.9\linewidth}}
    \toprule
    \textbf{\color{black}Code\color{black}} & \textbf{Description}\\
    \midrule
    10s1 & Left turn in busy intersection, navigating around object in road. \\
    10s2 & Slowing to avoid collision between two vehicles ahead of the ego vehicle. \\
    10s4 & Parallel parking between two cars on side of busy road. \\
    10s5 & Slowing to avoid being sideswiped during turn by a vehicle that crosses into ego vehicle's lane. \\
    10s6 & Sudden stop mid-intersection by large lead vehicle in adverse weather. \\
    10s7 & Left turn quickly followed by right merge for quick right turn in urban environment in adverse weather. \\
    10s8 & Ego vehicle slows for pedestrian crossing road in non-crosswalk area. \\
    4s1 & Reversing in parking lot. \\
    4s2 & Overtaking cyclist in suburban environment. \\
    4s3 & Ego vehicle must stop quickly for a pedestrian who jumps into the road. \\
    4s4 & Merging from far right lane to far left lane on highway to avoid emergency vehicles / car accident. \\
    4s5 & Ego vehicle must avoid object that falls off of lead vehicle on highway at night. \\
    4s6 & Ego vehicle hydroplanes on wet road on highway, and needs to maintain control. \\
    4s7 & Ego vehicle needs to pull over for flat tire at high speed. \\
    4s8 & Ego vehicle merges left to enter a busy highway. \\
    5s2 & Ego vehicle makes blind turn in intersection due to obstructed view. \textbf{[removed]} \\
    5s3 & Ego vehicle turning left and avoids a collision with another vehicle who ran a red light (T-bone). \\
    5s4 & Car in parallel lane merges into ego vehicle’s path. \\
    5s5 & Ego vehicle needs to navigate around a stopped cyclist. \textbf{[removed]} \\
    5s6 & Ego vehicle navigates construction zone at night. \\
    7s1 & Hidden stop sign at night. \\
    7s3 & Lead vehicle quickly decelerates (brake check) \\
    7s4 & Ego vehicle crosses the midline to overtake a slow lead vehicle. \\
    7s5 & Ego vehicle needs to slow quickly from high speed for a stopped cyclist around a turn. \textbf{[removed]} \\
    7s6 & Vehicle failure (flat tire) in small parking lot. \\
    xs2 &  Ego vehicle waits for child to cross crosswalk before turning right at stop sign. \\
    xs3 & Ego vehicle stops during right turn to avoid collision with vehicle turning right from incorrect (left) lane. \\
  \bottomrule
\end{tabular}
\end{table*}

We evaluated changes across four major outcomes of interest, with two additional descriptive outcomes, making a total of six ratings per scenario video. After each video, participants rated their: (1) comfort relying on the AV, (2) preference to take control, (3) satisfaction with the explanation, and (4) confidence in the AV's driving ability. We hypothesized that ratings may be context-dependent. As such, we also collected two ratings describing the driving context: (5) perceived harm and (6) perceived difficulty of driving in each scenario. These context descriptor variables were collected after the accurate explanation videos only. Outside of the rating task, participants answered questions about their trust in AVs, expertise, and demographics.

\subsection{Participants}
A total of 232 participants spread throughout the United States participated in the study and completed all study procedures. Participants were recruited from the general population via existing participant email lists and via SONA,\footnote{https://www.sona-systems.com} an undergraduate study pool system where students are granted study credit for participation. The mean age of the study sample was 23.6 (SD = 11.3), with ages ranging from 18 to 85 years. The sample was 73.7\% female.

\subsection{Simulated Driving Scenarios}
All simulated driving videos were custom-made by the research team using DReye VR~\cite{silvera2022dreye}, a tool for creating realistic driving scenarios using the open-source driving simulator CARLA~\cite{dosovitskiy2017carla}. Scenario design was inspired by the National Highway Traffic Safety Administration (NHTSA) list of common driving situations that result in vehicle crashes~\cite{najm2007pre}. Examples include unexpected pedestrians in the road, collision avoidance from erratic other drivers, construction and emergency vehicle zones, parallel parking and reversing from park, navigating around stopped or slowed vehicles or cyclists, and dealing with flat tires or hydroplaning. Videos lasted between 10 and 30 seconds each (Examples in Fig. \ref{scenarioexamples}). Three scenarios were removed during analysis due to the AV’s driving determined to be imperfect (going above the speed limit, crossing over the center line during a turn, and improper yielding), making a total of 24 included in the final analysis. Data was cleaned prior to analysis to ensure that all videos were viewed in full.

\subsection{AV Explanations and Errors}
During all driving scenario videos, explanations of the AV's behavior were provided to participants at the time of AV action, \color{black} modeled by the research team after those provided by~\citet{kim2018textual}. Explanations were presented both visually in written English on the vehicle's dashboard display and audibly via spoken English produced by Amazon Polly Text-To-Speech~\cite{polly}. These modalities are reflective of current trends in AV explanation research~\cite{kaufman2024effects, koo2015did, lee2023investigating}, providing accessibility to participants and ensuring that findings can be immediately incorporated into the design of state-of-the-art AVs and XAI interfaces. \color{black} For an example of the visual presentation of explanations, see Figure~\ref{scenarioexamples}.

Explanations provide information on `what' action the AV is doing (e.g. \textit{braking}) as well as a local explanation for `why' the vehicle is doing it (e.g. \textit{to avoid a collision with a merging vehicle}). The importance of `what' and `why'-type information has been studied in past experimental work on explanations of driving behavior by~\citet{kaufman2024effects} and~\citet{koo2015did}.  Frameworks by~\citet{wang2019designing} and~\citet{lim2019these} in cognitive science and~\citet{miller2019explanation} in philosophy emphasize the importance of `what' and `why' information for transparency, trust, and understanding with AI-based systems like AVs. We examine both the impact of `why' errors alone and `why and what' errors combined.

Explanation errors were presented via three conditions \color{black} (Table~\ref{conditions}). Those with an “accurate” explanation were correctly told `what' the AV was doing and `why' it was doing it. Errors were introduced via a “low” error condition, where the AV correctly explained `what' it was doing but incorrectly explained `why', and a “high” error condition, where both `what' and `why' explanations were incorrect. \color{black} Comparing the “accurate” to “low” group shows the impact of errors related to the AV's rationale for behavior (i.e. `why'). Comparing the “low” to “high” group isolates the impact of adding errors related to the AV's description of its own action (i.e. `what'). \color{black} We did not include a condition with inaccurate `what’ but accurate `why,’ as it would be implausible for an AV to provide a correct rationale for an incorrect action; this would reduce the study's ecological validity. \color{black}

\color{black}We hypothesized that users may process AV errors not just in terms of occurrence, but also in terms of potential outcome -- mirroring real-world scenarios where the consequences of an error are as critical as the error itself. \color{black} To test this, we conducted a secondary analysis, categorizing mistakes in the “high” error condition -- where the AV is providing an incorrect description of what it is doing -- based on the potential harm that \emph{would} result should the AV have acted on the mistaken `what' description. For example, if the AV mistakenly says it is going “left” when really it is going straight, we hypothesized that the impact of this error would be greater if going left would result in an accident, as opposed to simply a wrong turn. To explore if this difference in pragmatics does impact our results, we run a secondary analysis within just the 'high' condition to test for impact of \emph{potential} harm.

\begin{table*}[h] 
  \caption{Experimental Conditions. Participants saw videos across each \color{black} error \color{black} condition, providing ratings for each.}
  \Description{This table shows the three experimental conditions, providing a description and example of each. Participants saw videos across each condition, providing ratings for each.}
  \label{conditions}
    \vspace{-.5em}
  \begin{tabular}{p{0.1\linewidth}p{0.3\linewidth}p{0.36\linewidth}}
    \toprule
    \textbf{Condition} & \textbf{Explanation Description} & \textbf{Example}\\
    \midrule
    Accurate & Correct `what' and correct `why'& \textit{“Braking, a pedestrian is crossing the road.”} \\
    \midrule
    Low & Correct `what' and incorrect `why' & \textit{“Braking, a cyclist is crossing the road.”} \newline (When the obstacle is actually a pedestrian)\\
    \midrule
    High & Incorrect `what' and incorrect `why' & \textit{“Merging right, a cyclist is crossing the road.”} \newline (When the vehicle is slowing, not merging, and the obstacle is actually a pedestrian)\\
  \bottomrule
\end{tabular}
\end{table*}

\subsection{Measures}
\color{black} The measures in this study were carefully selected to offer clear and comprehensive insight into the impact of errors, context, and personal traits, while remaining easy and intuitive for participants to understand. \color{black}
\subsubsection{Main Outcomes: Scenario Ratings} 
These measures were taken after each video to provide insight into the impact of the explanation errors. \color{black} All measures were adapted from existing work and adjusted to fit the present study. \color{black}

\begin{itemize}[itemsep=5pt, topsep=0pt]
    \item \textbf{\underline{Comfort} relying on AV} (proxy for trust). “How comfortable would you feel relying on this AV in this specific situation?” \color{black} This was adapted from the Situational Trust Scale for Automated Driving (STS-AD) by \citet{holthausen2020situational}, which itself was based on Hoff and Bashir's trust in automation framework \cite{hoff2015trust}. \color{black} (0-10 rating scale)
    \item \textbf{\underline{Reliance} on AV} (preference to take control). “If this specific situation were to happen in the real world, would you prefer to rely on an AV or take control yourself?” \color{black} This was also adapted from the STS-AD scale \cite{holthausen2020situational}. \color{black} (Binary choice: 'Rely on AV' or 'Take control myself'). To aid comparison, reliance data was scaled from 0-1 to 0-10 to match the other variables. This scaling does not impact interpretation.
    \item \textbf{\underline{Satisfaction} with explanation.} “How satisfied are you with the AV’s explanation?” \color{black} This was adapted from the Explanation Satisfaction Scale by \citet{hoffman2018metrics}. \color{black} (0-10 rating scale)
    \item \textbf{\underline{Confidence} in AV driving ability.} “Please rate your confidence in the AV’s driving ability.” \color{black} This was adapted from the Performance Expectancy section of the Autonomous Vehicle Acceptance Model (AVAM) \cite{hewitt2019assessing}. \color{black} (0-10 rating scale). Note: the actual driving performance was always high and never changed between error conditions.
\end{itemize}

\subsubsection{Context Descriptors} 
These measures were taken for each scenario after the “accurate” condition video only to provide insight into the impact of driving context on main outcomes. \color{black} Both measures were based on the external variability factors impacting situational trust described by \citet{hoff2015trust}. \color{black}

\begin{itemize}[itemsep=5pt, topsep=0pt]
    \item \textbf{\underline{Harm} of Driving Situation.} “In the real world, how would you rate the risk of harm of this specific driving situation?” (0-10 rating scale)
    \item \textbf{\underline{Difficulty} of Driving.} “In the real world, how would you rate the difficulty of driving in this specific situation?” (0-10 rating scale)
\end{itemize}

\vspace{1em}
\subsubsection{Additional Outcomes}
These further contextualize our main experimental findings and were measured before and/or after the rating task.

\begin{itemize}[itemsep=5pt, topsep=0pt]
    \item \textbf{Expertise}. Expertise was measured before the rating task via three self-rated questions related to a person's knowledge and understanding of AVs, \color{black} adapted from \citet{kaufman2024effects}. \color{black} (5-point likert scale from Strongly Disagree to Strongly Agree)
    \item \textbf{Trust}. Trust was measured before and after the rating task via four questions related to adaptability, safety, overt trust, and willing to recommend a friend to ride in an AV. \color{black} Questions were summed to form a single, comprehensive composite measure. This is similar to the approach taken by \citet{hewitt2019assessing}, from whom these questions were adapted. \color{black}(5-point likert scale from Strongly Disagree to Strongly Agree)
    \item \textbf{Explicit Factors Contributing to Reliance Decisions}. After the task, participants rated the relative impact of seven aspects of the driving and explanation experience on their reliance decisions, \color{black} building an explicit understanding of which factors may be most important. \color{black}(5-point likert scale from Not At All to Very Much).
\end{itemize}

\begin{table*}[t] 
  \caption{Means and Standard Errors of Main Outcomes By Error Condition. As errors increased, outcome ratings decreased.}
  \Description{This table shows the Means and Standard Errors of each main outcomes, divided by Error Condition. As errors increased, outcome ratings decreased.}
  \label{summary_of_main}
      \vspace{-.5em}
  \renewcommand{\arraystretch}{1}
  \begin{tabular}{p{0.22\linewidth}p{0.1\linewidth}p{0.1\linewidth}p{0.1\linewidth}}
    \toprule
     & \textbf{Accurate} & \textbf{Low} & \textbf{High}\\
    \midrule
    Comfort Relying on AV & 4.59 (0.06) & 3.48 (0.06) & 2.71 (0.06) \\
    Reliance Decision & 3.65 (0.11) & 2.43 (0.10) & 1.60 (0.09) \\
    Satisfaction w/ Expl. & 6.38 (0.06) & 3.13 (0.06) & 2.20 (0.06) \\
    Confidence in Driving & 5.16 (0.06) & 3.68 (0.06) & 2.84 (0.06) \\
    \bottomrule
    \end{tabular}
\end{table*}

\section{Results}
\subsection{Impact of Explanation Errors: Comfort, Reliance, Satisfaction, and Confidence}
\subsubsection{Summary of Main Outcomes}
\begin{figure*}[t] 
  \centering
  \vspace{6em}
  \includegraphics[width=\linewidth]{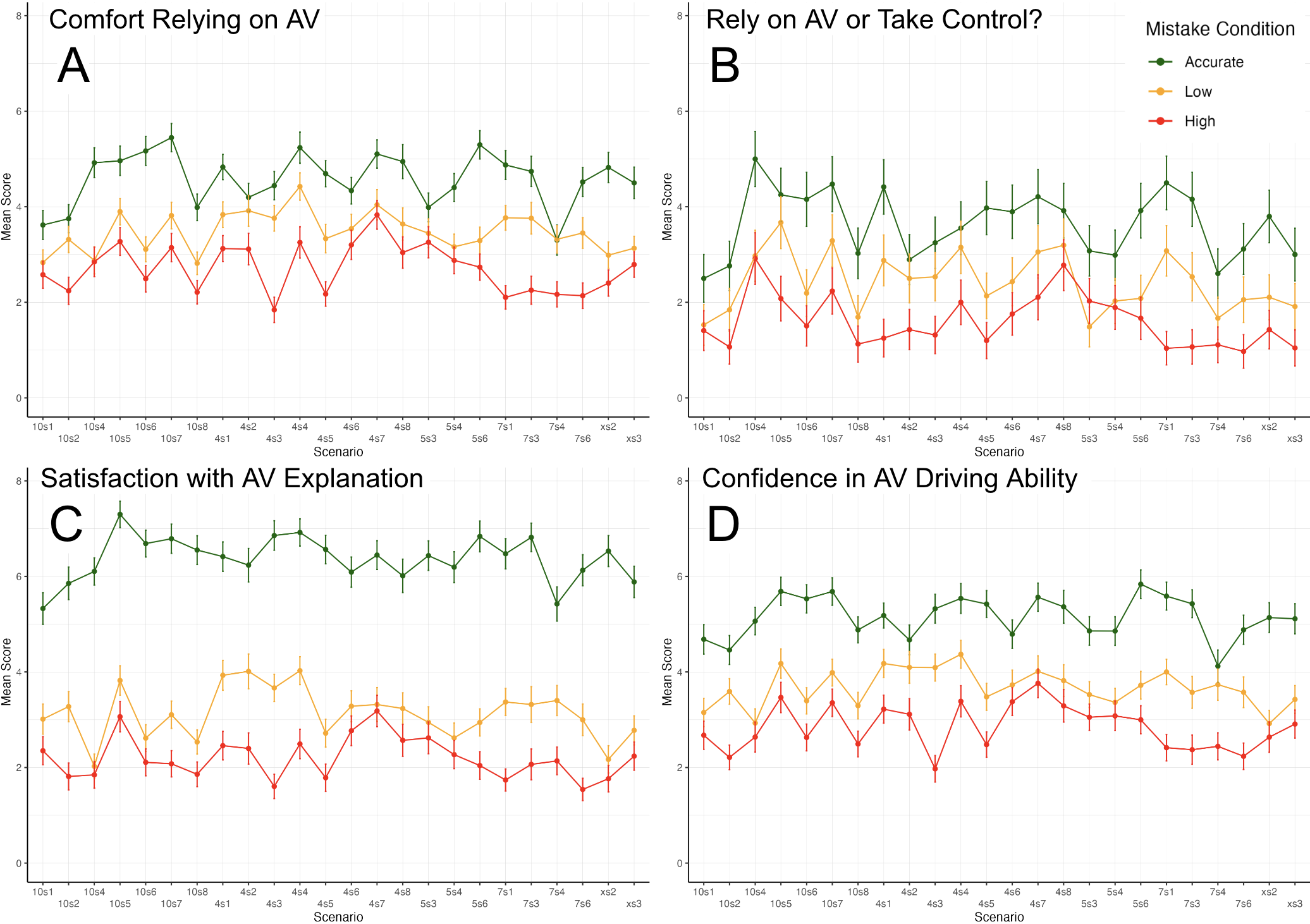}
  \vspace{-1em}
  \caption{Mean Outcome by Error Condition Across All Scenarios. Plot A shows mean Comfort Relying on AV, Plot B shows mean Reliance Preference (Binary, scaled to 1-10), Plot C shows mean Satisfaction with AV Explanation, and Plot D shows mean Confidence in AV Driving Ability. \color{black}Error bars are standard error. \color{black} Results consistently demonstrate Accurate > Low > High.}
  \Description{Four visualizations, one for each major outcomes by error condition, shown for all 24 scenarios, depicting the mean outcome score by error condition across. Plot A shows mean scores for Comfort Relying on AV, Plot B shows mean scores for Reliance Preference (Binary, scaled to 1-10), Plot C shows mean scores for Satisfaction with AV Explanation, and Plot D shows mean scores for Confidence in AV Driving Ability. Results consistently demonstrated the trend of Accurate > Low > High.}
  \label{individualscenarios}
\end{figure*}

Across our four main outcomes, segmenting the data by explanation error condition level gives us an initial impression on the impact of our experimental manipulation. We find the highest scores for comfort relying on the AV, reliance preference, satisfaction with the AV's explanation, and confidence in the AV's driving ability in the Accurate condition, followed by the Low condition and then the High condition (Table \ref{summary_of_main}). Visualizing main outcomes by scenario, we find that the effect of condition was very consistent across scenarios (Figure \ref{individualscenarios}\color{black})\color{black}.

Even when explanations were accurate, participants' overall comfort relying on the AV (comfort), and their preference to rely on the AV (reliance), were middling to low, reflecting the overall reluctance to trust AVs found in past human-AV interaction research  by~\citet{kenesei2022trust}. Participants were more positive about the AV explanations provided to them in the study, however, this satisfaction deteriorated quickly when errors were introduced. Despite the AV’s driving performance -- and therefore, demonstrated ability -- remaining consistent across all conditions, \emph{impressions} of the AV’s driving ability worsened as AV explanation errors were introduced.

\begin{table*}[t] 
  \caption{\color{black}Comparative effects of each Error Condition on Main Outcome variables (LME Models). \color{black} These show highly significant outcome score differences between each error level.}
  \Description{This table shows comparisons of each error condition for each main outcome (LME Models). These show highly significant outcome score differences between each error level.}
  \label{LMEoverall}
      \vspace{-1em}
  \renewcommand{\arraystretch}{1}
  \begin{tabular}{p{0.2\linewidth}p{0.05\linewidth}p{0.05\linewidth}p{0.05\linewidth}p{0.02\linewidth}|p{0.05\linewidth}p{0.05\linewidth}p{0.05\linewidth}p{0.02\linewidth}|p{0.05\linewidth}p{0.05\linewidth}p{0.05\linewidth}p{0.02\linewidth}}
    \toprule
    & \multicolumn{4}{c|}{\textbf{\small \color{black}Accurate (intercept) vs. Low}} & \multicolumn{4}{c|}{\textbf{\small \color{black}Accurate (intercept) vs. High}} & \multicolumn{4}{c}{\textbf{\small \color{black}Low (intercept) vs. High}} \\
    \midrule
    & \begin{math}\beta\end{math} & df & t & sig. & \begin{math}\beta\end{math} & df & t & sig. & \begin{math}\beta\end{math} & df & t & sig. \\
    \midrule
    Comfort Relying on AV & -1.08 & 5114 & -17.0 & *** & -1.85 & 5115 & -28.9 &  *** & -0.75 & 5110 & -11.9 &  ***\\
    Reliance Decision & -1.19 & 5114 & -10.2 & *** & -2.00 & 5115 & -17.1 & *** & -0.80 & 5108 & -6.8 & ***\\
    Satisfaction w/ Expl. & -3.24 & 5116 & -45.0 & *** & -4.16 & 5118 & -57.8 & *** & -0.93 & 5111 & -12.8 & ***\\
    Confidence in Driving & -1.46 & 5114 & -25.5 & *** & -2.29 & 5114 & -38.3 & *** & -0.83 & 5111 & -13.8 & ***\\
    \hline
    \multicolumn{13}{l}{\textit{Sig. Codes: ‘***’ p < 0.001 | ‘**’ p < 0.01 | ‘*’ p < 0.05 | ‘.’ trending p < 0.1 | ‘NS’ p $\geq$ 0.1} } \\
    \multicolumn{13}{l}{\color{black}\textit{Each outcome was modeled independently using the formula: $ \text{outcome} \sim \text{error\_level} + (1 \mid \text{scenario}) + (1 \mid \text{participant}) $} \color{black}}
\end{tabular}
\end{table*}

\subsubsection{Overall Impact of Errors on Main Outcomes}
Linear mixed-effects (LME) models were used to measure the effect of errors on each major outcome \color{black} independently\color{black}. Though LME models produce similar results to mixed-model ANOVAs, they offer greater flexibility for repeated measures experiments. By incorporating random effects, they reduce the likelihood of Type 1 errors~\cite{boisgontier2016anova}. \color{black} Separate models were made for each outcome, with fixed effects for error condition (Accurate, Low, High) \color{black} and random effects for the specific scenario and individual participant. The fixed effects allow us to compare differences between study conditions, while the random effects allow us to control for differences by scenario and individual participant \cite{meteyard2020best, barr2013random}. \color{black} An example formula is: \[\textit{$ \text{outcome} \sim \text{error\_level} + (1 \mid \text{scenario}) + (1 \mid \text{participant}) $}\]

The model was fit using restricted maximum likelihood (REML) with Satterthwaite’s approximation for degrees of freedom. In our LME models, the ‘Accurate’ condition was set as the baseline, with contrasts comparing ‘Accurate’ to ‘Low’ and ‘High’ conditions. To compare the ‘Low’ and ‘High’ conditions directly, the baseline was changed to ‘Low’. This contrast structure allowed us to compare the relative impact of each error level. Using treatment coding allows for direct comparisons between each error condition, and is preferred over distributing contrasts across all levels when the primary goal is to interpret each level's effect in terms of its deviation from a meaningful baseline \cite{meteyard2020best, barr2013random}. Direct contrasts were calculated within the LME model framework, eliminating the need for additional post-hoc comparisons. Following best practices for LME model reporting \cite{meteyard2020best} and recent related work using similar models \cite{kaufman2024effects}, we provide detailed estimates (\begin{math}\beta\end{math}), degrees of freedom (df), t-values (t), and p-values (sig.) for each fixed effect to enhance reproducibility. \color{black}

Individual comparisons between each condition (Accurate-Low, Accurate-High, Low-High) show highly significant effects (p < 0.001) across all four outcome measures (Figure \ref{LMEoverall}), implying error condition significantly impacted outcomes. The results for comfort relying on the AV, reliance decision, and satisfaction with the explanation follow the expected trend: we find that outcome scores decrease as error level increases. Surprisingly, we found the same effect on a person’s evaluation of the AV’s driving ability, where confidence scores also decrease as error level increases. This is unexpected, given that the driving performance shown in the videos were identical and only the explanations changed. The implication is that there is a cross-over effect between a person’s evaluation of the explanation and the person’s evaluation of the vehicle’s driving, \emph{despite} evidence suggesting that the driving performance is consistently high quality.

\subsubsection{Potential Harm of `What'-type Errors} 
In the high error condition group, what-type errors are incorrect descriptions of what the AV is doing. To understand the impact of the \emph{potential} harm of these errors, we conducted a secondary analysis comparing errors that would result in an accident if acted upon by the AV versus those that would not. \color{black} This analysis adds a more nuanced understanding for the impact of `what'-type errors on participant ratings of an AV. It is important to note that, though similarly named, this \textit{potential} harm categorization is unrelated to the perceived harm of the driving situation measured via participant rating. In the potential harm case, scenarios were categorized by the research team as harmful or not based on what the result would have been if the AV had driven in accordance with the mistaken `what' explanation. As the basis of our analysis, \color{black}  we use LME models with fixed effects for the categorization of potential harm (0 or 1) and random effects for driving scenario and individual differences by participant. This analysis is conducted only on data from the “high” condition in isolation.

We find significant effects for satisfaction with an explanation (\begin{math}\beta\end{math} = -0.46, t(22) = -2.6, p < .05) and confidence in the AV's driving ability (\begin{math}\beta\end{math} = -0.45, t(22) = -2.7, p < .05). Comfort relying on the AV showed results trending towards significance (\begin{math}\beta\end{math} = -0.38, t(22) = -1.9, p = 0.07). No significant results were found for the reliance preference measure. These results imply that the content of the error -- in this case, the potential harm that could result \emph{from} the error -- may impact the effect of the error on outcomes. Specifically, we find evidence that higher gravity errors may have a greater negative impact on some outcomes.

\subsection{Impact of Driving Context: Perceived Harm and Difficulty of Driving}
\subsubsection{Relationship Between Harm and Difficulty} 
\begin{figure*}[h] 
  \centering
  \includegraphics[width=.72\linewidth]{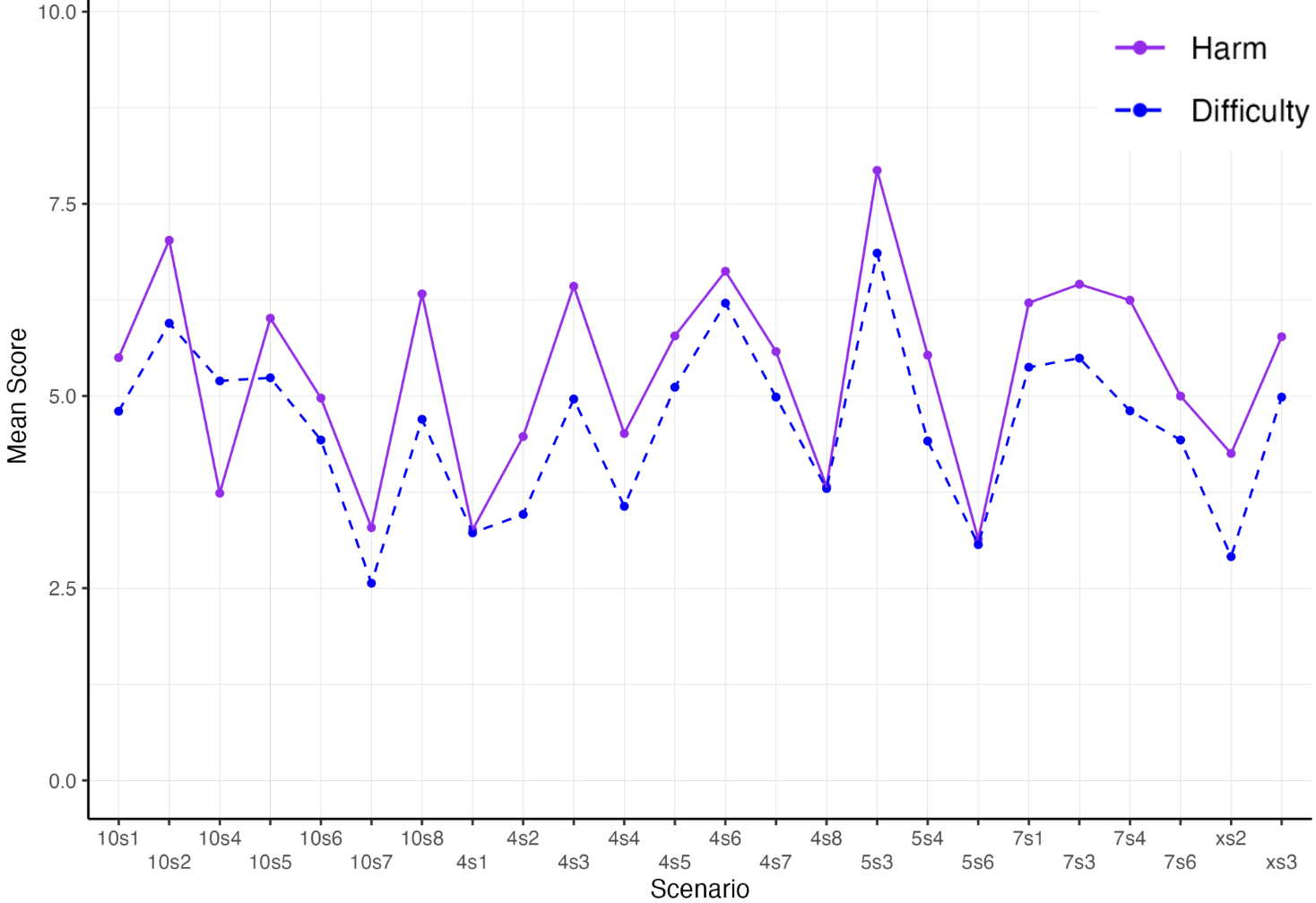}
  \vspace{-.5em}
  \caption{Mean Harm and Difficulty Across All Scenarios. Results show harm and difficulty are generally correlated, but differ depending on the specific scenario. We find that these contextual characteristics directly impacted outcome ratings as well as influenced the relationship between errors and outcomes.}
  \Description{A visualization of the mean Harm and Difficulty scores for all 24 scenarios, depicting the mean Harm and Difficulty ratings for each scenario. Results show harm and difficulty are generally correlated, but differ depending on the specific scenario. We find that these contextual characteristics directly impacted outcome ratings as well as influenced the relationship between errors and outcomes.}
  \label{individualdifficultyharm}
\end{figure*}

By examining the impact of harm and difficulty on outcome ratings, we can derive an understanding of how these contextual factors influenced our results. Unsurprisingly, we found a strong, positive relationship between the perceived harm and the difficulty of driving in a particular scenario (r = 0.69, p < .001; Adj R\textsuperscript{2} = .48). Figure \ref{individualdifficultyharm} shows difficulty and harm ratings by scenario.

\subsubsection{Impact of Harm and Difficulty On Main Outcome Ratings} 
We assess the impact of harm and difficulty on main outcome ratings using LME models with added fixed effects for the average harm and difficulty of scenarios. We maintain the fixed effect of \color{black}error \color{black} condition as well as random effects for scenario and participant \color{black} used previously, following the maximal structure described by \citet{barr2013random}. \color{black} Using this model, main effects for difficulty and harm give us an understanding of how these factors impacted outcome judgments \emph{independent} of error condition. \color{black} Once again, each outcome variable was modeled separately. \color{black} Estimates (\begin{math}\beta\end{math}) are given relative to the “accurate” condition. We find that harm significantly impacted comfort relying on the AV (\begin{math}\beta\end{math} = -0.43, t(38) = -3.7, p < .001), reliance decision (\begin{math}\beta\end{math} = -0.74, t(62) = -4.4, p < .001), and confidence in the AV's driving ability (\begin{math}\beta\end{math} = -0.22, t(41) = -2.2, p < .05). No significant effects were found between harm and explanation satisfaction. We found that difficulty impacted only reliance decision (\begin{math}\beta\end{math} = 0.60, t(61) = 3.0, p < .01). Interestingly, this was in the opposite direction as harm, where greater difficulty was associated with increased reliance. 

The implication is that contextual factors like harm and difficulty affect interaction with an AV system regardless of error condition, with harm potentially being the more influential factor. Specifically, higher harm is generally associated with lower comfort, reliance, and confidence, while higher difficulty is generally associated with higher reliance ratings. There does not appear to be an overall association between harm or difficulty and explanation satisfaction.

\begin{table*}[!htbp] 
  \caption{Impact of Harm on Error Condition Differences  \color{black} for each Main Outcome variable \color{black} (LME Models). These show if Harm influences the \textit{impact} of an error. We find significant effects for several outcomes, implying that Harm may moderate how much of an impact an error may have.}
  \Description{This table shows the impact of Harm on Error Condition Differences (LME Models). These show if Harm influences the impact of an error. We find significant effects for several of our outcomes, implying that Harm may moderate how much of an impact an error may have.}
  \label{harminteraction}
      \vspace{-.5em}
  \begin{tabular}{p{0.28\linewidth}p{0.05\linewidth}p{0.05\linewidth}p{0.05\linewidth}p{0.05\linewidth}|p{0.05\linewidth}p{0.05\linewidth}p{0.05\linewidth}p{0.05\linewidth}}
    \toprule
    & \multicolumn{4}{c|}{\textbf{\color{black}Accurate (intercept) vs. Low}} & \multicolumn{4}{c}{\textbf{\color{black}Accurate (intercept) vs. High}} \\
    \midrule
    & \begin{math}\beta\end{math} & df & t & sig. & \begin{math}\beta\end{math} & df & t & sig. \\
    \midrule
    Comfort Relying on AV & 0.43 & 5108 & 4.2 & *** & 0.10 & 5108 & 0.9 &  NS \\
    Reliance Decision & 0.35 & 5106 & 1.8 & . & 0.19 & 5106 & 1.0 & NS \\
    Satisfaction w/ Expl. & 0.13 & 5108 & 1.1 & NS & -0.15 & 5108 & -1.3 & NS \\
    Confidence in Driving & 0.27 & 5108 & 2.8 & ** & -0.01 & 5108 & -0.1 & NS  \\
    \hline
    \multicolumn{9}{l}{\textit{Sig. Codes: ‘***’ p < 0.001 | ‘**’ p < 0.01 | ‘*’ p < 0.05 | ‘.’ trending p < 0.1 | ‘NS’ p $\geq$ 0.1}} 
\end{tabular}
\end{table*}

\begin{table*}[!htbp] 
  \caption{Impact of Difficulty on Error Condition Differences \color{black} for each Main Outcome variable \color{black} (LME Models). These show if Difficulty influences the \textit{impact} of an error. We find significant effects for several outcomes, implying that Difficulty may moderate how much of an impact an error may have.}
  \Description{This table shows the impact of Difficulty on Error Condition Differences (LME Models). These show if Difficulty influences the impact of an error. We find significant effects for several of our outcomes, implying that Difficulty may moderate how much of an impact an error may have.}
  \label{difficultyinteraction}
      \vspace{-.5em}
    \begin{tabular}{p{0.28\linewidth}p{0.05\linewidth}p{0.05\linewidth}p{0.05\linewidth}p{0.05\linewidth}|p{0.05\linewidth}p{0.05\linewidth}p{0.05\linewidth}p{0.05\linewidth}}
    \toprule
    & \multicolumn{4}{c|}{\textbf{\color{black}Accurate (intercept) vs. Low}} & \multicolumn{4}{c}{\textbf{\color{black}Accurate (intercept) vs. High}} \\
    \midrule
    & \begin{math}\beta\end{math} & df & t & sig. & \begin{math}\beta\end{math} & df & t & sig. \\
    \midrule
    Comfort Relying on AV & -0.26 & 5109 & -2.1 & * & 0.1 & 5109 & 0.7 &  NS \\
    Reliance Decision & -0.42 & 5107 & -1.9 & . & -0.16 & 5107 & -0.7 & NS  \\
    Satisfaction w/ Expl. & -0.08 & 5109 & -0.6 & NS & 0.30 & 5109 & 2.1 & * \\
    Confidence in Driving & -0.21 & 5108 & -1.9 & . & 0.04 & 5109 & 0.3 & NS \\
    \hline
    \multicolumn{9}{l}{\textit{Sig. Codes: ‘***’ p < 0.001 | ‘**’ p < 0.01 | ‘*’ p < 0.05 | ‘.’ trending p < 0.1 | ‘NS’ p $\geq$ 0.1}}
\end{tabular}
\end{table*}

\subsubsection{Impact Of Harm and Difficulty On Differences Between Error Conditions (Interaction Effects)} 
Using similar LME models \color{black} as used to assess the main effect of error conditions, \color{black} we can assess if difficulty or harm impacted the relationship \emph{between} error level and outcomes by looking at interaction effects. Specifically, we examine if the \emph{differences} found between each error condition are predicted by a scenario's average difficulty or harm \color{black}(for example, for harm, we can modify our formula to look at the interaction between level and harm by looking at the effect of \textit{error\_level * mean\_harm} \cite{barr2013random})\color{black}. In this case, we use the accurate group as a common reference (model intercept) and then compare if the changes of each main outcome from accurate to low and accurate to high vary with respect to difficulty and harm. \color{black}Just as before, separate models were made for each outcome. \color{black} We find significant interaction effects for several of our outcomes, implying that contextual characteristics like harm and difficulty may be moderating how much of an impact an error may have. Table \ref{harminteraction} shows interaction effects for harm, and Table \ref{difficultyinteraction} shows interaction effects for difficulty.

\textbf{Comfort Relying on AV} -- For comfort relying on the AV, we find that scenario difficulty has more of an impact on comfort in the low error condition than in the accurate condition. Specifically, when difficulty increased, comfort decreased significantly more in the low condition compared to the accurate condition. A similar but opposite effect was found with harm: in the low error condition, comfort increases significantly more with higher harm than in the accurate condition. The implication is that comfort judgments may be more sensitive to the difficulty and harm of the scenario when there are some errors (low condition) compared to when there are no errors (accurate). We do not see difficulty or harm as more impactful on comfort judgments in the high error condition compared to the accurate condition. This implies that these judgments of comfort are likely based on the high magnitude of error (main effect) for the high group, as opposed to being influenced by the difficulty or harm of the driving context (interaction effect) in these cases. 

\textbf{Reliance Decision} -- The case is similar for reliance decision, though the effects are trending towards significance as opposed to being statistically significant. We find that scenario difficulty has more of an impact on reliance in the low error condition than it did in the accurate condition. Specifically, when difficulty increased, reliance decreased more in the low condition compared to the accurate condition. For harm in the low error condition, higher harm increases reliance significantly more than in the accurate condition. The implication is that reliance judgments may be more sensitive to the difficulty and harm of the scenario when there are some errors (low condition) compared to when there are no errors (accurate). As with comfort, we do not find difficulty or harm as more impactful on reliance in the high error condition compared to the accurate condition. This implies that judgments of reliance when errors are high are likely based on the error itself rather than difficulty or harm.

\textbf{Satisfaction w/ Expl.} -- The interaction effects change for satisfaction with an explanation. We do not see difficulty or harm as more impactful on satisfaction in the low error condition compared to the accurate condition, implying that differences in judgments of the explanation are likely based on the explanation quality itself without being influenced by the difficulty or harm of the situation in these cases. The same effect is seen when comparing differences between the accurate and high condition groups for harm. Surprisingly, we find that difficulty differentially impacted explanation satisfaction when in the high condition compared to the accurate condition. Specifically, when difficulty increased, satisfaction increased significantly more when in the high condition than when in the accurate condition. These results together imply that, in most cases, judgments of an explanation are likely based on the explanation quality itself, however, when explanation quality is exceptionally poor and driving is difficult, participants may actually be more forgiving with their judgments.

\textbf{Confidence in Driving} -- Lastly, for confidence in the AV's driving ability, we find a similar trend as comfort and reliance decision. We find that scenario difficulty has more of an impact on confidence in the low error condition than it did in the accurate condition. When difficulty increased, confidence decreased more in the low condition compared to the accurate condition (trending towards significance). For harm in the low error condition, higher harm increases confidence significantly more than in the accurate condition. The implication is that confidence judgments may be more sensitive to the difficulty and harm of the scenario when there are some errors (low condition) compared to when there are no errors (accurate). As with comfort and reliance, we do not find difficulty or harm as more impactful on confidence in the high error condition compared to the accurate condition. This implies that judgments of confidence are more likely based on the main effect of error itself, rather than difficulty or harm.

\begin{figure*}[t] 
  \centering
  \includegraphics[width=0.72\linewidth]{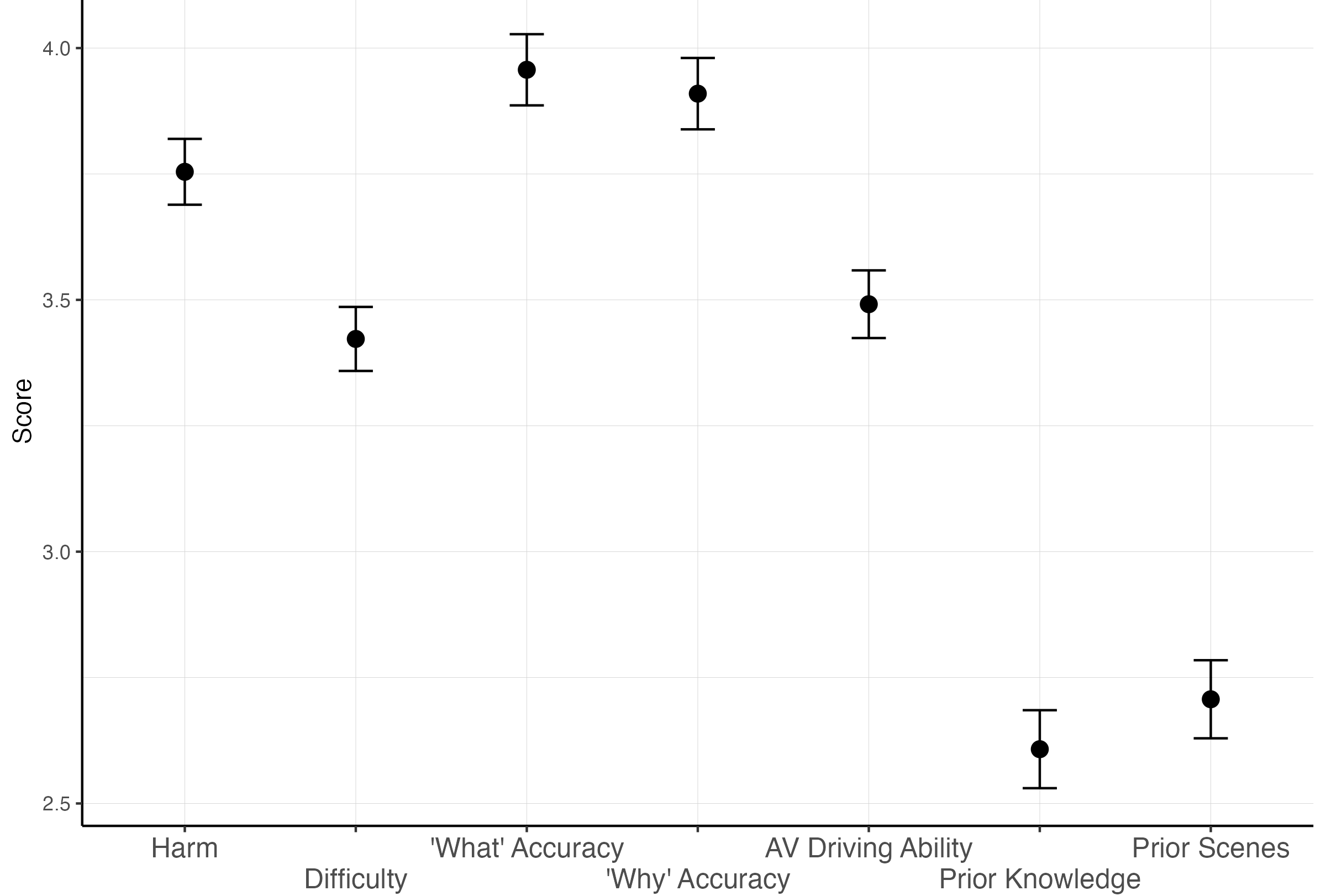}
  \vspace{-0.5em}
  \caption{Relative Importance of Factors on Reliance Decisions. This shows the mean \color{black}self-reported \color{black} importance rating of each factor with respect to a person's reliance decision \color{black} (errors bars are standard error). \color{black} Results show \color{black} the accuracy of each type of information, perceived contextual harm and difficulty, and the driving ability of the AV were all important factors\color{black}.}
  \Description{A visualization of relative importance (mean rating) of seven factors on reliance decisions. Results show many of the factors of interest in this study were of high importance.}
  \label{factors}
\end{figure*}

\subsection{Trust and Expertise}
\subsubsection{Impact Of Exposure To Errors On Trust}
We measured participant trust before and after the experiment was completed to see if exposure to repeated AV explanation mistakes affected AV trust levels as a disposition. In general, using a paired-samples t-test, we did not find that exposure to errors in the experiment impacted trust. Only one result was significant pre-post experiment: participants felt that they “understand why an autonomous vehicle makes decisions” worse after the experiment as compared to before (p < 0.001).

\subsubsection{Correlations of Initial Trust and Subjective Expertise with Outcomes}
Calculating correlations between initial trust and self-reported expertise level and a participants’ average outcome ratings allows us to assess if trust or expertise can predict how a person will evaluate the AV. We found a strong positive correlation between initial trust and expertise, (r = 0.58, p < 0.001), implying that those who know more about AVs trust them more.

We found moderate positive correlations between expertise and comfort (r = 0.38, p < 0.001), satisfaction (r = 0.35, p < 0.001), and confidence (r = 0.36, p < 0.001), meaning that those with more expertise reported more positively to these outcomes on average. We found high correlations between initial trust and comfort (r = 0.52, p < 0.001) as well as confidence (r = 0.51, p < 0.001). We found moderate correlations between initial trust and reliance (r = 0.30, p < 0.001) as well as satisfaction (r = 0.49, p < 0.001). These effects are consistent with linear mixed-effects models looking at the main effects of expertise and initial trust on outcomes. Together, these results imply that participants with higher initial trust or expertise tended to rate outcomes higher on average.

\subsection{Self-Reported Rationale for Reliance Decisions}
\subsubsection{Factors Contributing to Rating Decisions}
We explicitly inquired upon the basis of reliance decisions by asking participants to rate the relative importance of several factors on their reliance ratings (Figure \ref{factors}). \color{black} This analysis helps clarify which elements participants explicitly prioritized, offering valuable guidance for designing explanations that align with user preferences and expectations. \color{black} Ratings were given after the main scenario experiment concluded. As a whole, we found that the accuracy of the explanation on ‘what’ the AV is doing was the most important consideration \color{black}(mean = 3.96, SE = 0.07)\color{black}, followed by the explanation accuracy of ‘why’ the AV is doing it \color{black}(mean = 3.91, SE = 0.07)\color{black}, the harm of the situation \color{black}(mean = 3.75, SE = 0.07)\color{black}, the AV’s driving ability \color{black}(mean = 3.49, SE = 0.07)\color{black}, and the difficulty of the driving situation \color{black}(mean = 3.42, SE = 0.06)\color{black}. Prior scenes viewed \color{black}(mean = 2.71, SE = 0.08) \color{black} and prior knowledge \color{black}(mean = 2.61, SE = 0.08) \color{black} did not have much reported impact, though this is unsurprising, as these implicit judgments may be difficult for participants to self-describe. Many factor differences were significant based on \color{black}Kruskal-Wallis test ($\chi^{2}$ = 285.7, df = 6, p < 0.001). For pairwise comparisons, we use Dunn's test and adjust p-values using Bonferroni's method. \color{black} Notably, we find harm was significantly more important than difficulty \color{black}(p < 0.05)\color{black}, while explanation accuracy (both `what' and `why' components) were significantly more important than AV driving ability \color{black}(both p < 0.001). \color{black} This latter effect may be in part due to the salience of these factors based on the study's manipulation. We did not see significant differences between the importance of `what' and `why' accuracy.

\subsubsection{Feedback from Participants}
Participants were given the opportunity to express their general views on autonomous vehicles on how explanation errors impacted their decision-making specifically. Unsurprisingly, many participants reflected general concerns with AVs, including doubt that they can \textit{“adapt to any circumstance … [causing] a bigger accident.”} One participant commented that \textit{“after a lifetime of driving myself, I'm not sure if I feel comfortable giving over control to a computer.”} Desire for control was brought up multiple times. The common sentiment was that giving control to AVs is considered risky. 

Regarding exposure to explanation errors, one participant commented that they were \textit{“less confident in the reliability of [autonomous] vehicles”} after exposure to errors, while another commented that when explanations \textit{“were wrong, [it made] my confidence in the system shaky.”} These comments reiterate our findings on the negative impact of errors on reliance, as well as the holistic judgment of an AV's ability in general. Some participants broke their decision-making down in terms of individual factor priority. For instance, \textit{“if the AV described the action it was taking incorrectly I was a lot less confident and consistently chose to take control myself … if the reason was incorrect, it still impacted my confidence level but not as much.”} This reflects the general trend found in our individual factor analysis suggesting `what' information may be more important than `why' rationale, even if this trend was not found to be statically significant.

Regarding the impact of context, a participant noted, \textit{“I chose to take control myself in higher risk situations, even if the AV did a good job navigating.”} This supports this study's overall findings on the importance of contextual factors like harm on decision-making.

\color{black}
\section{Discussion}
This study aimed to assess the impact of autonomous vehicle (AV) explanation errors and driving context characteristics on participant comfort relying on an AV, preference to rely on the AV instead of taking control themselves, satisfaction with the explanation, and confidence in the AV’s driving ability. Using a mixed-methods approach with a heterogeneous sample of participants (n = 232), we found that explanation errors and contextual characteristics like driving difficulty and perceived harm have a large impact on how a person may think, feel, and behave towards AVs. Echoing prior work by~\citet{nourani2020role}, important consideration must also be given to personal factors like initial trust and expertise, which may further impact how a person interacts with the system. Our results provide insight into how to design effective human-AV interactions and interactions with AI-based systems more generally. We discuss key findings and their implications for future AV design in these contexts.

\vspace{0.5em}
\textbf{Autonomous vehicle (AV) explanation errors had a detrimental effect on all outcomes \color{black}(RQ1)\color{black}.} Our results suggest that errors significantly reduced participant comfort relying on an AV, preference to rely on the AV instead of taking control themselves, and satisfaction with the explanation. These effects are unsurprising, as we would expect a person's reliance decision or satisfaction with an explanation to reflect the system's performance~\cite{hoff2015trust}. 

We were surprised, however, to find crossover effects between the \emph{explanatory} performance of the AV and a person's confidence in the AV’s \emph{driving ability}, given the AV’s actual demonstrated driving performance remained consistent across all conditions. This crossover effect alludes to the mental model of potential AV users, where evaluation of the explanatory performance of the vehicle and the driving performance of the vehicle are connected. Though connecting explanatory and driving performance may be an intuitive and correct assumption, empirically observing this crossover provides evidence supporting the importance of high quality explanatory interfaces, particularly within safety-critical systems. \color{black}Further highlighting this point, the finding that explanation accuracy was explicitly rated as more important than driving ability for reliance decisions underscores the role of transparency and comprehensibility in user acceptance. This highlights the need for AV systems to prioritize clear and accurate explanations of actions (`what’) and rationales (`why’) in building trust, and suggests that users may evaluate AV systems as much by their communications as by their actual driving performance. \color{black} Taken together, in the case of AVs, this means that -- even if the AV's driving performance is perfect -- if the explanations produced by the AV are not accurate as well, people may still refuse to adopt AV technology. This insight likely generalizes to explanatory communications for other AI-based systems.

\color{black}Even with accurate explanations, Comfort, Reliance Preference, and Confidence in the AV's driving ability were nowhere near ceiling level -- hovering near the middle of the provided scale. Though the middling scores may be reflective of the overall challenging nature of the driving situations presented in the study, even the most straightforward scenarios did not have a mean reliance preference score greater than mid-range. These results reflect the trend found in a plethora of prior work supporting the premise that people do not generally trust or wish to adopt autonomous vehicles, reemphasizing this as an important area of continued study. \color{black}

\vspace{0.5em}
\textbf{The negative impact of errors increased with error magnitude and the potential harm of the error \color{black}(RQ1 continued)\color{black}.} Expanding on the effect of errors observed in our study, we find that the impact was commensurate with the magnitude of the errors presented. In the study presented here, ‘what' and `why’ errors (high condition) in combination had worse outcomes than ‘why’-only errors (low condition). On the surface, this is unsurprising, as more errors demonstrates lower system performance, and these results could simply be the cumulative effects of seeing errors on two parts of the explanation instead of just one. 

There remains a high probability that `what' and `why' information errors play distinct roles in the evaluation of the system's performance, however. This would align with past work showing how description of action and description of justification may differentially impact the way a person interacts with autonomous systems~\cite{koo2015did, kaufman2024effects, miller2019explanation}. When asked explicitly about the factors that contributed to their reliance decisions, participants reported that explanation accuracy (both `what' and `why' components) were both important. Qualitative assessment of participant feedback, however, suggests that ‘what’ information may be more important for reliance outcomes than `why' rationale. This may be due to the potential harm that improper action (`what') can have on driving safety, in contrast to improper justification (`why'). Concretely, a car incorrectly turning right into traffic can cause physical harm, but incorrectly justifying \emph{why} it is turning right can not. \color{black} This finding highlights that explanation accuracy, particularly for ‘what’ information, is crucial not only for user comfort but also for the perception of the AV’s reliability and overall system competence. \color{black} We note that \emph{conclusively} differentiating the impact of `what' from ‘why’ errors independently will be left for future work.

Supporting the hypothesis that the \emph{implications} of an error matter in addition to the mere \emph{presence} of an error, we find that ‘what’ errors -- with more dire implications for potential harm -- had a greater negative impact on comfort, confidence, and explanation satisfaction. Specifically, in cases when `what' errors would have resulted in vehicle crashes if the AV had acted upon them, people were far less comfortable relying on the AV, had less confidence in the AV's driving, and preferred the explanation less. We posit that the only reason we did not see differences in reliance preference as well was because the presence of an error altogether already reduced reliance preference to near-minimum levels. 

Together, these results suggest that the implications of an explanation error in terms of what they may mean about the system's performance and consequences of harm, affect how a person thinks, feels, and behaves with an AV. Implications may be implicitly calculated based on the amount and type of error, as well as how these intersect with the external driving context.

\vspace{0.5em}
\textbf{Contextual factors like the difficulty of driving in a specific situation and the perceived harm of that situation directly affect how people think, feel, and prefer to behave with AVs \color{black}(RQ2)\color{black}.} We found evidence that driving difficulty and perceived harm directly influenced outcomes regardless of error condition. Specifically, higher harm was associated with lower comfort, reliance, and confidence ratings, while higher difficulty was associated with higher reliance ratings. We did not find direct associations between harm or difficulty and explanation satisfaction.

We attribute the negative impact of harm to the common concern that AVs may malfunction, and a malfunction in a more harmful driving situation may have more severe implications. This reflects the general sentiment that many people think \emph{they} can drive better than an AV in most situations. The seemingly positive impact of difficulty on reliance may reflect that, in very high difficulty situations, people lack the confidence that they could perform better than the AV and, as such, prefer to rely on the AV in these cases. Indeed, the driving demonstrated by the AV in even the most difficult driving situations presented in the study was high quality. 

Explicitly, participants reported contextual harm as significantly more important than driving difficulty for their reliance decision. \color{black}Theoretically, this may be because difficult driving situations often entail a higher risk of harm, especially when participants are uncertain about the AV's ability to handle these challenges. We hypothesize that reliance decisions are grounded in an overall evaluation of potential harm, with driving difficulty acting as one indicator of this risk. Accordingly, \color{black} difficulty and AV driving performance matter \emph{because} harm matters, making harm itself the underlying driver of reliance decisions. 

A person's decision to rely on an AV (and other outcomes) may be a function of the demonstrated performance of the vehicle -- compounded by an implicit evaluation of the overall harm of the situation (in part as a function of difficulty) -- weighted against that person's confidence in their own ability to perform better. This is similar to the model proposed by~\citet{hoff2015trust}, with added details on the nested relationship between driving difficulty and harm.

The general implication of these findings is that there is a complex relationship between contextual factors like harm and difficulty, and outcomes like comfort, reliance, and confidence. It is clear that models of AV behavior incorporating internal and external characteristics of the driving situation can bring further insight into how people will interact with AVs~\cite{kaufman2024developing}, and future research should examine how these can be best supported by explanatory communication. 

\vspace{0.5em}
\textbf{Contextual factors moderate the relationship between errors and outcomes in complex, sometimes counter-intuitive ways \color{black}(RQ2 continued)\color{black}.} By examining the interaction between contextual characteristics and the effects of error conditions, we found that -- in some cases -- the driving context may influence the \emph{amount} of impact an error has. The general trend we observed for comfort relying on the AV, reliance preference, and confidence in the AV's driving ability was that when difficulty increased, outcome scores decreased \emph{more} in the low error condition compared to the accurate condition, with no differences seen between the accurate and high conditions. The same but opposite effect was found for harm: higher harm increased outcomes effect in the low condition more than the accurate condition, with no difference between the accurate and high condition.

The lack of difference found in the way difficulty and harm moderate the relationship between the accurate and high conditions implies that differences in these judgments are likely based solely on the error level as opposed to difficulty or harm of the driving context. Concretely, the effect of error likely overrides the effect of context when errors are at the extremes. When errors are in the middle -- such as for the low error condition -- the impact of the error appears to be more strongly moderated by the context. This may be because the ramifications of `why'-type errors are not strong enough to drive effects fully based on the presence of the error, and instead the amount of change that the error can cause is based on how dire the implications may be in a particular context. The direction of effect, in this case, is similar to the main effects explained previously.

For explanation satisfaction, we do not see difficulty or harm as more impactful on satisfaction in the low error condition, if we compare this to the accurate condition. This implies that differences in satisfaction are likely based on the explanation quality itself in these cases. The same effect is seen between the accurate and high condition for harm, but not for difficulty. Surprisingly, we found that when difficulty increased, satisfaction increased significantly more when in the high condition than when in the accurate condition. Together, these results imply that, in most cases, judgments of an explanation are based primarily on the explanation quality itself, however, when explanation quality is especially poor and driving is difficult, participants may actually be more forgiving of the AV.

Taken together, results from this analysis on interaction effects illustrate a nuanced and often complex relationship between driving difficulty, perceived harm and error condition when predicting our main outcomes. This nuance alludes to a prioritization of how different aspects of the driving situation -- including harm, difficulty, and explanation errors -- combine to form judgments of comfort in the AV, reliance preference, satisfaction with an explanation, and confidence in an AV's driving ability.

\vspace{0.5em}
\textbf{Trust and expertise influenced how people think, feel, and behave towards the AV \color{black}(RQ3)\color{black}.} As discussed, how a person thinks and behaves towards AVs is not just a function of their external environment. We find evidence that internal characteristics, such as a participant’s trust and expertise, may also have played a role in the study’s results. Participants with higher initial trust or expertise tended to rate study outcomes higher on average. We also found that those with higher AV expertise trusted AVs more in general. As with the function on driving reliance described previously, these findings also align with Hoff and Bashir's model of trust with AI systems, where a person's prior experience impacts their reliance with the system~\cite{hoff2015trust}. Those with higher expertise may also have other underlying characteristics, like an affinity for new technologies, which may have influenced their trust. We attribute the finding that priors were not reported as explicitly important for reliance decisions in our post-evaluation to the common difficulty articulating the impact of implicit dispositions.

Though trust may be influential, we did not find evidence that exposure to AV mistakes due to the study’s manipulation had a large effect on a person’s overall trust level. A single exception was that participants reported less understanding of why AVs make decisions once the study concluded. This makes sense given the inconsistent explanations for behavior they had received, which may have impacted a person's mental model of AV decision-making.

\subsection{Implications For Autonomous Vehicle Design and Research}
Our results have important implications for future AV design and research \color{black}\textbf{(RQ4)}\color{black}. Understanding the consequences of AV explanation errors and contextual characteristics like driving difficulty and perceived harm are necessary first steps towards designing vehicles which may be more trustworthy, reliable, and satisfactory for people interacting with the AV. \color{black} By using current, multimodal XAI methods of explanation presentation -- visual and auditory cues of both `what' and `why' information -- this study’s findings can directly apply to current AV explanation design. \color{black} In this section, we will discuss how our study's insights can be applied to build more trustworthy AVs.

The foremost implication of this work is to emphasize the importance of designing systems that can produce accurate explanations for AV decisions. This implication is not too surprising, as why would designers ever \emph{intentionally} produce inaccurate explanations? It becomes more meaningful, however, when testing explanatory systems before deployment. Our results indicate that even with a well-functioning driving system, if explanations are not of high quality, people still won't want to use the AV.

The crossover between a person's evaluation of \emph{explanation} performance and \emph{driving} performance may be of particular interest to AV designers. If this crossover is a concern -- as it should be in the case of deployment -- the most obvious solution would be to work on improving explanations so that they are on par with driving ability. In the meantime, specific UI features may be implemented to help users separate their evaluations of explanations from driving performance, such as by presenting distinct indicators for each system reassuring a rider that even if the explanation is messed up, the car can still drive sufficiently. This may be challenging or misleading, however, particularly if the explanatory system and driving system are indeed connected.

Though not tested in the present study, our results also have implications for \emph{conditionally} automated vehicles -- those which may require human takeover in difficult or computationally complex driving situations (SAE Level 3)~\cite{inagaki2019critique, ayoub2021investigation}. There is the possibility that explanation errors produced in conditional contexts may result in inappropriate behaviors by human passengers. If users lose trust in the AV due to explanation errors, for example, they may react unpredictably or take over control at inappropriate times, potentially leading to accidents.

For both the fully autonomous and conditionally autonomous cases, our results indicate that explanatory systems should be deployed with confidence in their accuracy or potentially not at all. Future work can estimate the exact outcome differences between no explanation and inaccurate explanations, but combining the results of our study and prior findings on the importance of AI explainability by~\citet{gunning2019darpa} and~\citet{miller2019explanation} suggests that \emph{accurate} explanations should be opted for whenever possible, and deploying untested or inconsistent systems can be disastrous.

In terms of design priority, our work suggests that if explanation designers are going to focus on improving a single aspect of the explanation, they should optimize for accurate `what' information before `why' justification. Though both may be necessary in the long run, providing a correct description of behavior may produce a higher value return on effort for teams looking to find a place to start. Just as they have for differentiating the effects of accurate `what' and `why' explanations of driving behavior, future work should delineate the potentially different impacts `why' and `what' \emph{errors} may have from each other in isolation. This can allow more detailed and conclusive conclusions on their relative importance to outcomes such as those in the present study.

It is important to note that, while explanation satisfaction scores were generally higher than other outcome scores in the accurate condition, explanations were still nowhere near ceiling level. This means that there are likely ways in which our explanations could have improved. Prior work has emphasized outcome differences based on informational content, modality (such as including visualizations or haptic feedback), timing, or interactivity~\cite{miller2019explanation, kaufman2024effects, avetisyan2022investigating, goldman2022trusting}. In this regard, future AV research and design teams should continue iterating on these aspects of explanation design and human-machine interface (HMI) development in order to fine-tune explainable systems.

Another major implication of this work is to orient AV designers to the potentially different outcomes that may result from interactions by people with different traits or while operating in driving environments. \color{black}In AVs, user familiarity with driving norms allows for immediate identification of errors, which intensifies the effect on trust and reliance. This contrasts with fields where users cannot as easily detect errors without domain expertise, highlighting the importance of context-specific trust calibration mechanisms for XAI systems. Specific to the case studied here, \color{black} context-aware or personalized explanations may be necessary in cases where the driving difficulty or perceived harm are classified as greater, or for people with little expertise or initial trust. It is likely that designers will need to focus on limitations on cognitive resources like attention and load, particularly for high difficulty scenarios. \color{black} In high-risk domains like AVs, these factors may be particularly important, as the potential consequences of explanation errors are immediate and tangible. We thus contribute to an emerging understanding of how explanation accuracy and error detectability shape user perceptions in safety-critical environments. \color{black} Other personal traits like risk preferences or personality may also be a basis for personalization~\cite{bockle2021can}. Segmenting populations or isolating particular driving conditions to test future explanation designs would be an effective method to figure out how to meet the differing needs of diverse user groups~\cite{kaufman2024developing}.

Taken in full, our results provide insight into the impact of explanation errors, and provide direction for future AV researchers and designers to improve human-AV communication. Given the serious negative repercussions that may result from errors, our results also provide a foundation for the creation of ethical or regulatory guidelines which dictate the testing and accuracy of AV explanations before they can be deployed to real-world consumers.

\subsection{Limitations and Future Study}
As with any study, this research is not without limitations. First, though study findings were generally robust and fit within logical narratives, there is the possibility that findings resulting from online data collection and based on simulations may not generalize to real-world attitudes or behavior. A large effort was put into making driving simulations as realistic as possible, however, we could not test the effect of explanation errors on real roads out of concern for participant safety. Future work should seek to test the impact of errors in a real-world environment. A similar concern may be found for study outcomes: though results on reported reliance or comfort may be clear in a research context, there is no guarantee that explicit declarations of thought or behavior will remain consistent when immersed in a real-world driving context with real consequences of bodily or financial harm. This is a concern for all studies which rely on survey measures or explicit participant statements as a proxy for real-world behavior. It is possible that seeing each scenario three times (with different explanations) may have impacted the ratings provided. This could be mitigated in the future by showing participants only one video per scenario, randomized by error condition. Finally, we examined proximal explanations for action (\textit{“braking”}) and cause of action (\textit{“… a pedestrian is crossing the road.”}) presented in written fashion and auditory fashion. Future work can examine explanations of distal causes (“pedestrian in the road … because there is an obstacle on the sidewalk”) which could provide additional context for \emph{why} a driving situation is happening in the first place. These could be explained using potentially different modalities of presentation.

\section{Conclusion}
In a simulated driving study with 232 participants we tested how autonomous vehicle (AV) explanation errors, driving context characteristics (perceived harm and driving difficulty), and personal traits (prior trust and expertise) impact four driving perception and behavior-related outcomes: comfort relying on an AV, preference to rely on the AV instead of taking control themselves, satisfaction with the explanation, and confidence in the AV’s driving ability. 

Our results indicate that explanation errors, contextual characteristics, and personal traits have a large impact on how a person may think, feel, and behave towards AVs. Explanation errors negatively affected all outcomes. Surprisingly, this included reduced ratings of the AV's driving ability, despite driving performance remaining constant. The negative impact of errors increased with error magnitude and the potential harm of the error, providing evidence that \emph{implications} of an error matter in addition to the mere \emph{presence} of an error. Harm and driving difficulty directly impacted outcomes as well as moderated the relationship between errors and outcomes. In general, harm was associated with lower comfort, reliance, and confidence ratings, while driving difficulty was associated with higher reliance ratings. Overall harm was the more important contextual factor of consideration. In terms of a decision function for AV reliance, perceived harm -- influenced by driving difficulty -- may underlie the evaluation between a person's confidence in the AV's performance compared to their own ability. We found that individuals with higher expertise tended to trust AVs more, and these each correlated with more positive outcome ratings in turn. 

Overall, our results emphasize the need for accurate and contextually adaptive AV explanations to foster trust, reliance, satisfaction, and confidence. Understanding the ramifications of explanation errors can help future AV research and design teams prioritize design and better understand the impacts of their design choices. They also provide a foundation for context-aware design, personalized explanation interfaces, and potential ethical or regulatory guidelines for the deployment of explainable AI (XAI) systems for autonomous vehicles that are safe and trustworthy.



\begin{acks}
We would like to thank Saumitra Sapre, Rohan Bhide, Chloe Lee, Janzen Molina, and Emi Lee for driving scenario development and simulator testing.
\end{acks}

\newpage
\bibliographystyle{ACM-Reference-Format}
\bibliography{bibliography}

\end{document}